\documentclass{JHEP3}

\usepackage{epsfig,multicol,bbm}
\usepackage{amsmath}
\usepackage{amsfonts}
\usepackage{amssymb}
\usepackage{url}

\newcommand{\di}{{\rm d}}
\newcommand{\eqreff}[1]{Eq.~\eqref{#1}}
\newcommand{\ztr}{z_{\rm tr}}

\title{The extreme limit of the generalised Chaplygin gas}

\author{Oliver F. Piattella\\
Dipartimento di Scienze Fisiche e Matematiche,
Universit\`a dell'Insubria,
Via Valleggio 11, 22100 Como, Italy\\
and\\
INFN, sezione di Milano, Via Celoria 16, 20133 Milano, Italy\\
and\\
Institute of Cosmology and Gravitation, University of Portsmouth,
Dennis Sciama Building, Portsmouth PO1 3FX, United Kingdom\\
E-mail: \email{oliver.piattella@uninsubria.it}}

\preprint{\arXivid{0906.4430}}

\abstract{Unified Dark Matter models describe Dark Matter and Dark Energy as a single entity which is, in the simplest case, embodied in a perfect barotropic fluid. It is a well-established fact that small adiabatic perturbations of Unified Dark Matter have an evolution characterised by oscillations and decay which provide predictions on the Cosmic Background Radiation anisotropies which are in poor agreement with observation. In this paper we investigate the generalised Chaplygin gas and we find that the Integrated Sachs-Wolfe effect excludes the model for $10^{-3} < \alpha < 350$. We discuss the implications on the background evolution of the Universe if large values of $\alpha$ are considered. In this case, the Universe expansion mimics a matter-dominated phase abruptly followed by a de~Sitter one at the transition redshift $\ztr$. Thanks to an analysis of the type Ia supernovae Constitution set we are able to place $\ztr = 0.22$.}


\begin{document}

\section{Introduction}

The generalised Chaplygin gas \cite{Kamenshchik:2001cp} (hereafter gCg) is a cosmological model which has raised some interest thanks to its ability to describe alone the dynamics of the entire dark sector of the Universe, i.e. Dark Matter and Dark Energy. The gCg is a perfect fluid characterised by the following equation of state:
\begin{equation}\label{gcgeos}
p  = - \frac{A}{\rho^{\alpha}}\;,
\end{equation}
where $p$ is the pressure, $\rho$ is the density and $A$ and $\alpha$ are positive parameters. 

In the setting of the Friedmann-Lema\^{i}tre-Robertson-Walker cosmological theory, one easily integrates the energy conservation equation and finds that the gCg density evolves as a function of the redshift $z$ as follows:
\begin{equation}\label{gcgrhoevo}
\rho = \left[A + B\left(1 + z\right)^{3(1 + \alpha)}\right]^{\frac{1}{1 + \alpha}}\;,
\end{equation}
where $B$ is a positive integration constant. Equation~\eqref{gcgrhoevo} interpolates between a past ($z \gg 1$) dust-like phase of evolution, i.e. $\rho \sim \left(1 + z\right)^{3}$, and a recent ($z \lesssim 1$) one in which $\rho$ tends asymptotically to a cosmological constant, i.e. $\rho \sim A^{\frac{1}{1 + \alpha}}$.

As cosmological model, the gCg was first investigated in \cite{Kamenshchik:2001cp} because it had already raised some attention, e.g. in connection with string theory. In particular, the equation of state (\ref{gcgeos}) can be extracted from the Nambu-Goto action for $d$-branes moving in a $(d + 2)$-dimensional spacetime in the lightcone parametrisation, see \cite{Bordemann:1993ep}. Also, the gCg is the only fluid which, up to now, admits a supersymmetric generalisation, see \cite{Hoppe:1993gz, Jackiw:2000cc}.  

From the point of view of cosmology, a possible unification picture between Dark Matter and Dark Energy is particularly appealing, especially in connection with the so-called cosmic coincidence problem (see \cite{Zlatev:1998tr} for much detail about the latter). This motivation prompted an intensive study of the gCg and of those models that have the property of unifying (dynamically) Dark Matter and Dark Energy. Such models are called Unified Dark Matter (UDM).

It must be pointed out that in a conventional cosmological model (i.e. a model in which Dark Matter and Dark Energy are separated entities) Dark Matter has the primary role of providing for structure formation whereas Dark Energy has to account for the recently observed accelerated expansion of the Universe \cite{Riess:1998cb, Perlmutter:1998np}. The fundamental task for the gCg is then to play both these roles.

To understand if this is the case, the gCg parameters space $(A,\alpha)$ has been intensively analysed in relation with observation of Large Scale Structures (LSS), Cosmic Microwave Background (CMB), type Ia Supernovae (SNIa), X-ray cluster gas fraction and Baryon Acoustic Oscillations (BAO), see for example \cite{Wu:2006pe, Davis:2007na, Lu:2008hp, Lu:2008zzb, Barreiro:2008pn, Sandvik:2002jz, Carturan:2002si,Bean:2003ae, Amendola:2003bz, Amendola:2005rk, Giannantonio:2006ij}. 

The most instructive constraints are about the parameter $\alpha$ and come from the LSS matter power spectrum and the CMB angular power spectrum analysis. In the former, the resulting best fit value is $\alpha \lesssim 10^{-5}$, \cite{Sandvik:2002jz}. One can think that this narrow constraint is a flaw of the gCg model because if we take the limit $\alpha \to 0$ in the equation of state \eqreff{gcgeos}, then we reproduce exactly the $\Lambda$CDM model (if $A$ assumes the value of the cosmological constant energy density).

The authors of \cite{Beca:2003an} have found a loophole in this degeneracy problem. Indeed, they have shown that if we consider a baryon component added to the gCg and we compute the matter power spectrum for the baryonic part alone, it turns out that the latter is only poorly affected (at the linear perturbative regime) by the presence of the gCg. The model constituted by gCg plus baryons is therefore in good agreement with LSS observation for all $\alpha \in (0,1)$. 

On the other hand, even if the ``baryon loophole'' allows to circumvent the very narrow constraint on $\alpha$ coming from LSS analysis, it cannot prevent the tight one coming from CMB analysis and due in particular to the Integrated Sachs-Wolfe (ISW) effect, \cite{Carturan:2002si,Bean:2003ae, Amendola:2003bz, Amendola:2005rk, Giannantonio:2006ij, Bertacca:2007cv}. In more detail, if we take into account an ordinary Cold Dark Matter (CDM) component and a baryonic one together with the gCg, we find that $\alpha < 0.2$, \cite{Amendola:2003bz}. Withdrawing the CDM component, we lower the bound by an order of magnitude: $\alpha < 10^{-2}$, \cite{Amendola:2005rk}. Finally, for the case of the pure gCg we find an even tighter constraint: $\alpha < 10^{-4}$, \cite{Bertacca:2007cv}. 

Taking into account all these results, we may conclude that the gCg model is viable only when it is very similar, if not degenerate, to the $\Lambda$CDM. However, it must be pointed out that in the major part of the literature on the subject, the preliminary constraint $\alpha < 1$ is assumed. The reason for this assumption resides in the form of the gCg adiabatic speed of sound, which is the following:
\begin{equation}\label{gCgsos}
 c_{\rm s}^{2} \equiv \frac{\di p}{\di\rho} = \frac{A\alpha}{A + B\left(1 + z\right)^{3\left(\alpha + 1\right)}}\;.
\end{equation}
If $z \to -1$ then $c_{\rm s}^{2} \to \alpha$. Therefore, causality would require $\alpha < 1$. However, the causality problem for UDM models should rather be addressed in terms of a microscopic theory, see \cite{Babichev:2007dw}. In the particular case of the gCg, for $\alpha > 1$ the authors of \cite{Gorini:2007ta} examine in some detail the causality issues and develop a suitable microscopic theory in which the signal velocity never exceeds the speed of light.

Given this, it is natural to wonder how the gCg model behaves in the ``forbidden'' range $\alpha > 1$. In the literature few papers carry out such analysis, for example \cite{Amendola:2005rk, Gorini:2007ta, Fabris:2008hy, Fabris:2008mi, Urakawa:2009jb}. An interesting result is for instance that, in the framework of the ``baryon loophole'' above mentioned, i.e. considering the model constituted by the gCg plus baryons, the agreement of the baryon power spectrum with observation increases for $\alpha \gtrsim 3$, see \cite{Gorini:2007ta}. In \cite{Fabris:2008hy} and \cite{Fabris:2008mi} the authors have recently confirmed this behaviour. On the other hand, CMB constraints do not change significantly from the case $\alpha < 1$, in the sense that the parameter is again constrained to be very small, see \cite{Amendola:2005rk} and \cite{Urakawa:2009jb}.

Even in those papers which analyse the gCg for $\alpha > 1$, the parameter space is probed up to a maximum finite value. For example, $\alpha < 6$ in \cite{Amendola:2005rk} or $\alpha < 10$ in \cite{Urakawa:2009jb}. Since the $\alpha > 1$ case has proved to be promising, at least in relation with LSS analysis, in this paper we dedicate ourselves to investigate the extreme limit of the gCg, i.e. the behaviour of the model for very large values of $\alpha$. Our analysis is principally based on the ISW effect because, as we have discussed in the above and as the authors of \cite{Bertacca:2007cv} have shown, it provides the strongest constraints for a UDM model.

To this purpose, in section II we briefly outline the basic equations describing the ISW effect and, inspired by \cite{Bertacca:2007cv}, we present a simple method based on the M\'esz\'aros effect which we employ in Section III to find a qualitative constraint for large values of $\alpha$. Indeed, in section III we analyse the Jeans wavenumber of gCg perturbations and we find that if $\alpha \gtrsim 350$ the ISW effect can be potentially small. In Section IV we then confirm the results of Sec. III by directly calculating the ISW effect in the limit $\alpha \to \infty$ and showing that it is smaller than the one for $\alpha = 0$, i.e. for the $\Lambda$CDM (the calculation of the ISW effect for the $\Lambda$CDM has been performed for the first time by Kofman and Starobinsky in 1985, see \cite{Kofman:1985fp}). In section V we address the behaviour of the background expansion for large values of $\alpha$ and find that the evolution is characterised by an early dust-like phase abruptly interrupted by a de~Sitter (dS) one. We then analyse the 157 nearby SNIa of the Constitution set and find that the transition between the two phases takes place about a redshift $z_{\rm tr} = 0.22$, which is much more recent than the transition at $z_{\rm tr} = 0.79$ to the accelerated phase of expansion in the $\Lambda$CDM model. The last section is devoted to discussion and conclusions.

\section{The ISW effect and the constraining method}

In this paper we discuss only adiabatic perturbations and we assume a flat spatial geometry. The ISW effect contribution to the CMB angular power spectrum is given by the following formula \cite{Sachs:1967er}:
\begin{equation}\label{ClISW}
 \frac{2l + 1}{4\pi}C_{l}^{\rm ISW} = \frac{1}{2\pi^{2}}\int_{0}^{\infty}\frac{\di k}{k}k^{3}\frac{\left|\Theta_{l}\left(\eta_{0},k\right)\right|^{2}}{2l + 1}\;,
\end{equation}
where $l$ is the multipole moment, $k$ is the wavelength number, $\eta_{0}$ is the present conformal time and 
\begin{equation}\label{ThetalISW}
 \frac{\Theta^{\rm ISW}_{l}\left(\eta_{0},k\right)}{2l + 1} = 2\int_{\eta_{*}}^{\eta_{0}}\Phi'\left(\eta,k\right){\rm j}_{l}\left[k\left(\eta_{0} - \eta\right)\right]\di\eta\;
\end{equation}
is the fractional temperature perturbation generated by the ISW effect, where $\eta_{*}$ is the last scattering conformal time, $\Phi\left(\eta,k\right)$ is the Fourier transformed gravitational potential and ${\rm j}_{l}$ is the spherical Bessel function of order $l$. The prime denotes derivation with respect to the conformal time $\eta$. For a more detailed description of the perturbations equations and of the integrals (\ref{ClISW}-\ref{ThetalISW}) we refer the reader to \cite{Sachs:1967er, Bardeen:1980kt, Mukhanov:1990me, Hu:1995em}.

Following \cite{Mukhanov:1990me}, consider the Fourier transformed evolution equation for the gravitational potential:
\begin{equation}\label{equ}
u'' + k^{2}c_{\rm s}^{2}u - \frac{\theta''}{\theta}u = 0\;,
\end{equation}
where:
\begin{equation}
u \equiv \frac{2\Phi}{\sqrt{\rho + p}}\;, \qquad\mbox{ and }\qquad \theta \equiv \sqrt{\frac{\rho}{3(\rho + p)}}(1 + z)\;,
\end{equation}
and $c_{\rm s}$, $\rho$ and $p$ are, respectively, the adiabatic speed of sound, the energy density and the pressure defined for a generic cosmological model. We have chosen here units such that $8\pi G = c = 1$. For simplicity, here and in the following we do not explicit the $\left(\eta,k\right)$ dependence for $u$ and $\Phi$ and the $\eta$ dependence for $c_{\rm s}^{2}$, $\theta$, $\rho$, $p$ and $z$. They will therefore be always implied, unless otherwise stated.

In \eqreff{equ} define 
\begin{equation}\label{kJ2def}
 k^{2}_{\rm J} \equiv \frac{\theta''}{c_{\rm s}^{2}\theta}
\end{equation}
as the square Jeans wavenumber. In general, when wavelengths smaller than the Jeans scale enter the Hubble horizon, they start to oscillate, affecting the CMB power spectrum and the matter one in ways not compatible with observation. For UDM models in particular, the authors of \cite{Bertacca:2007cv} have found a contribution to the ISW effect proportional to the fourth power of the speed of sound which generates in the CMB angular power spectrum a growth proportional to $l^{3}$ until $l \approx 25$ (the value $l \approx 25$ is related to the equivalence wave number $k_{\rm eq}$ which we discuss later). See \cite{Bertacca:2007cv} and also \cite{Hu:1995em} for more detail. 

Consequently, if we take into account only those scales for which 
\begin{equation}\label{cond}
 k^{2} < k_{\rm J}^{2}\;,
\end{equation}
then the gravitational potential does not oscillate.

Of course if $c_{\rm s}^{2} = 0$, as for the $\Lambda$CDM model, the Jeans scale vanishes and condition (\ref{cond}) is satisfied for every $k$ at any time (remember that $k_{\rm J}^{2}$ is time dependent). On the other hand, this is not the only possible scenario. In fact, the cosmological scales which are important for the CMB and structure formation are those which entered the Hubble horizon after the matter-radiation equivalence epoch. Those which entered the horizon earlier had been damped by the dominating presence of radiation. This effect is known as M\'esz\'aros effect \cite{Hu:1995em, Meszaros:1974tb, Weinberg:2002kg, Coles:1995bd}.

If we require that the relevant scales, i.e. $k < k_{\rm eq}$, must satisfy condition (\ref{cond}) we obtain:
\begin{equation}\label{cond2}
 k^{2}_{\rm eq} < \frac{\theta''}{c_{\rm s}^{2}\theta}\;.
\end{equation}
This relation can be used to infer qualitative constraints upon a generic cosmological model. In the next section we will make use of it to find constraints on $\alpha$. As a remark, a scenario for which \eqreff{cond2} holds true without demanding a vanishing speed of sound is the so-called {\it fast transition}, introduced and investigated in great detail in \cite{Piattella:2009kt}.

\section{Constraints on the generalised Chaplygin gas}

The equivalence wavenumber $k_{\rm eq}$ has the following form:
\begin{equation}
 k_{\rm eq}^{2} = \frac{H_{\rm eq}^{2}}{c^2\left(1 + z_{\rm eq}\right)^{2}}\;,
\end{equation}
where $H_{\rm eq}$ is the Hubble parameter evaluated at the equivalence redshift $z_{\rm eq}$. From the 5-year WMAP observation\footnote{\url{http://lambda.gsfc.nasa.gov/}}, the best fit values are $z_{\rm eq} = 3176^{+151}_{-150}$ and $k_{\rm eq} = 0.00968 \pm 0.00046$ $h$ Mpc$^{-1}$, where $h$ is the Hubble constant in 100 km s$^{-1}$ Mpc$^{-1}$ units. See also \cite{Komatsu:2008hk, Dunkley:2008ie}.

The Hubble parameter is related to the Universe energy content by Friedmann equation, which for the pure gCg model has the following form:
\begin{equation}\label{gcgH}
 \frac{H^{2}}{H_{0}^{2}} = \left[\bar{A} + \left(1 - \bar{A}\right)\left(1 + z\right)^{3(\alpha + 1)}\right]^{\frac{1}{\alpha + 1}}\;,
\end{equation}
where $\bar{A} \equiv A/(A + B)$ and $H_0$ is the Hubble constant. Notice that $\bar{A} = -w_0$, where $w_0$ is the present time equation of state parameter of the Universe.

Let $z_{\rm tr}$ be the redshift at which the accelerated phase of expansion begins. From \eqreff{gcgH}, we calculate the following relation between $\bar{A}$ and $z_{\rm tr}$:
\begin{equation}\label{Abaralpharel}
 \bar{A} = \frac{\left(1 + z_{\rm tr}\right)^{3\left(\alpha + 1\right)}}{2 + \left(1 + z_{\rm tr}\right)^{3\left(\alpha + 1\right)}}\;.
\end{equation}
From now on we make use of $\left(z_{\rm tr},\alpha\right)$ as independent parameters.

\begin{figure}
\begin{center}
  \includegraphics[width=0.7\columnwidth]{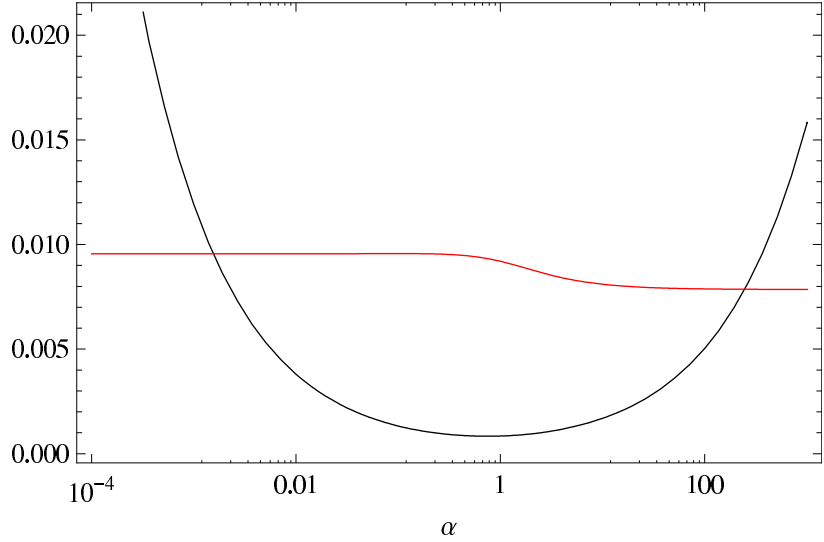}
  \caption{$k_{\rm J}$ (black curve) and $k_{\rm eq}$ (red ``quasi-horizontal'' line) as functions of $\alpha$. $k_{\rm J}$ is evaluated at $z = 0$ and $z_{\rm tr} = 0.79$, while $k_{\rm eq}$ is evaluated at $z_{\rm eq} = 3176$. The wavenumbers are in units $h$ Mpc$^{-1}$.}
\label{Fig1}
\end{center}
\end{figure}

Plugging Eqs. (\ref{gcgeos}), (\ref{gcgrhoevo}), (\ref{gCgsos}), (\ref{gcgH}) and (\ref{Abaralpharel}) into the definition (\ref{kJ2def}), we plot in Fig.~\ref{Fig1} the Jeans wavenumber as function of $\alpha$ for fixed $z = 0$ and $z_{\rm tr} = 0.79$ and the equivalence wavenumber, computed from \eqreff{gcgH}, as function of $\alpha$ for a fixed $z_{\rm eq} = 3176$. The value we have chosen for the transition redshift $z_{\rm tr}$ is the WMAP5 best fit, see \cite{Komatsu:2008hk, Dunkley:2008ie}.

The most intriguing feature of Fig.~\ref{Fig1} is that the Jeans wavenumber has a minimum value for $\alpha \approx 1$. For sufficiently small or sufficiently large values of $\alpha$ it grows and equals $k_{\rm eq}$. From our numerical computation we have inferred the following constraints: $\alpha \lesssim 10^{-3}$ and $\alpha \gtrsim 250$.

It is also possible to obtain these results from approximations, but analytically. Indeed, when $\alpha \ll 1$ we expand in Taylor series $k_{\rm eq}^{2}$ and $\alpha k_{\rm J}^{2}$ and find:
\begin{eqnarray}
 k^{2}_{\rm eq} &=& \frac{\left(1 + z_{\rm tr}\right)^{3} + 2\left(1 + z_{\rm eq}\right)^{3}}{\left(1 + z_{\rm eq}\right)^{2}\left[2 + \left(1 + z_{\rm tr}\right)^{3}\right]} + O(\alpha)\;,\\ \nonumber\\
\alpha k_{\rm J}^{2} &=& \frac{3\left[4 + \left(1 + z_{\rm tr}\right)^{3}\right]}{4\left(1 + z_{\rm tr}\right)^{3}} + O(\alpha)\;.
\end{eqnarray}
To leading order in $\alpha$, equating the above expressions gives the upper bound for small values of $\alpha$. For $z_{\rm eq} = 3176$ and $z_{\rm tr} = 0.79$: $\alpha \lesssim 10^{-3}$. 

Now expand $k_{\rm eq}^{2}$ in Taylor series for $\alpha \gg 1$:
\begin{equation}\label{keqalphainf}
 k^{2}_{\rm eq} = \frac{1 + z_{\rm eq}}{\left(1 + z_{\rm tr}\right)^{3}} + O\left(\frac{1}{\alpha}\right)\;.
\end{equation}
The corresponding expansion for $k_{\rm J}^{2}$ is less immediate. Making use of the asymptotic forms of the speed of sound and of the Hubble parameter, which we will give in \eqreff{cs2alphainf} and in \eqreff{gcgH2alphainf}, we find:
\begin{equation}\label{kJ2alphainf}
k_{\rm J}^{2} = \left\{
\begin{array}{cl}
x^{3\alpha}\left[\dfrac{6}{\alpha}\dfrac{x^4}{\left(1 + z_{\rm tr}\right)^2} + O\left(\dfrac{1}{\alpha^2}\right)\right]     & \mbox{   for   } x > 1\\ \\
\dfrac{9\alpha}{4}\dfrac{1}{\left(1 + z\right)^2}\left[1 + O\left(x^{3\alpha}\right)\right] & \mbox{   for   } x < 1
\end{array}
\right.\;,
\end{equation} 
where we have defined
\begin{equation}
 x \equiv \frac{1 + z}{1 + z_{\rm tr}}\;.
\end{equation}
From \eqreff{kJ2alphainf}, to leading order in $1/\alpha$ the Jeans wavenumber is an exponential function of $\alpha$ for $x > 1$; for $x < 1$, the expansion can be performed only with respect to $x^{3\alpha}$ and, to leading order, $k_{\rm J}^{2}$ grows linearly with $\alpha$. We take into account the latter instance, equate (\ref{kJ2alphainf}) to (\ref{keqalphainf}) (each expression considered to its leading order) and find $\alpha \gtrsim 250$ (for $z = 0$, $z_{\rm eq} = 3176$ and $z_{\rm tr} = 0.79$).

Note that we have neglected the CDM, the baryon and the radiation components. Neglecting the CDM component is reasonable, because the gCg model aims to an unified description of Dark Matter and Dark Energy. Adding a CDM component would then spoil its purpose. Moreover, neglecting radiation is also reasonable, because the minimum value of the Jeans wavenumber is at late times ($z \approx 0$), where radiation is subdominant. For what concerns baryons, their presence would have the effect of lessening the average speed of sound, increasing thus the Jeans wavenumber and smoothing the constraints we have found. Nonetheless, at late times the baryon component is also subdominant with respect to the gCg one, so we can reasonably neglect it.

On the other hand, in the calculation of $k_{\rm eq}^{2}$ we are not allowed to neglect both radiation and baryons, because at $z_{\rm eq} = 3176$ they are important. Therefore, since at early times the gCg and the $\Lambda$CDM model are indistinguishable, we use the WMAP5 result $k_{\rm eq}~=~0.00968$~$h$~Mpc$^{-1}$ and we find that $\alpha \lesssim 10^{-3}$ and $\alpha \gtrsim 350$. As we expected, taking into account the radiation contribution has the effect of increasing the lower bound for large values of $\alpha$.

\section{Calculation of the ISW effect for the extreme limit of the generalised Chaplygin gas}

In this section we calculate the integrals (\ref{ClISW}) and (\ref{ThetalISW}) in the limit $\alpha \to \infty$. To this purpose, in place of \eqreff{equ}, we employ the evolution equation for the gravitational potential $\Phi$, namely
\begin{equation}\label{eqPhi}
\Phi'' + 3\mathcal{H}\left(1 + c_{\rm s}^{2}\right)\Phi' + \left(2\mathcal{H}' +
\mathcal{H}^{2} + 3\mathcal{H}^{2}c_{\rm s}^{2} + k^{2}c_{\rm s}^{2}\right)\Phi = 0\;,
\end{equation}
where $c_{\rm s}^2$ is the gCg adiabatic speed of sound defined in (\ref{gCgsos}) and $\mathcal{H} = a'/a$ is the Hubble parameter written in the conformal time (see \cite{Mukhanov:1990me} for more detail). In the limit of very large values of $\alpha$, from \eqreff{gCgsos} together with \eqreff{Abaralpharel}, the square speed of sound has the following asymptotic behaviour
\begin{equation}\label{cs2alphainf}
c_{\rm s}^{2} = \left\{
\begin{array}{cl}
\dfrac{\alpha}{2}\left[x^{-3\alpha} + O\left(x^{-6\alpha}\right)\right]   & \mbox{   for   } x > 1\\  \\
\alpha\left[1 - 2x^{3\alpha} + O\left(x^{6\alpha}\right)\right] & \mbox{   for   } x < 1
\end{array}
\right.\;,
\end{equation}
with $c_{\rm s}^{2} = \alpha/3$ for $x = 1$. Friedmann equation (\ref{gcgH}) becomes:
\begin{equation}\label{gcgH2alphainf}
\frac{H^{2}}{H_0^2} = \left\{
\begin{array}{cl}
x^{3}\left[1 + \dfrac{\ln2}{\alpha} + O\left(\dfrac{1}{\alpha^2}\right)\right] & \mbox{   for   } x > 1\\ \\
1 + \dfrac{2x^{3\alpha}}{\alpha} + O\left(\dfrac{x^{6\alpha}}{\alpha^2}\right) & \mbox{   for   } x < 1
\end{array}
\right.\;,
\end{equation}
with $\tfrac{H^{2}}{H_0^2} = 1 + \tfrac{\ln3}{\alpha} + O\left(\tfrac{1}{\alpha^2}\right)$ for $x = 1$. To leading order in $1/\alpha$, the solution of \eqreff{gcgH2alphainf} for $x > 1$ as a function of the conformal time has the following form:
\begin{equation}\label{gcgH2alphainfsol1}
a = \frac{1}{1 + z_{\rm tr}}\left(\frac{\eta}{\eta_{\rm tr}}\right)^2\;,
\end{equation}
where $\eta_{\rm tr}$ is the conformal time corresponding to the transition redshift $z_{\rm tr}$. For $x < 1$, let $\eta_0$ be the present epoch conformal time and normalise the scale factor as $a(\eta_0) = 1$; the corresponding solution of \eqreff{gcgH2alphainf} is:
\begin{equation}\label{gcgH2alphainfsol2}
a = \frac{1}{1 + \eta_0 - \eta}\;.
\end{equation}
Joining solutions (\ref{gcgH2alphainfsol1}) and (\ref{gcgH2alphainfsol2}) in $a(\eta_{\rm tr})$, we can link the transition and the present epoch conformal time to the transition redshift as follows:
\begin{equation}
\eta_0 - \eta_{\rm tr} = z_{\rm tr}\;.
\end{equation}
Note that the relevant contribution to the ISW effect comes only from solution (\ref{gcgH2alphainfsol2}). In fact: $i)$ the background solution (\ref{gcgH2alphainfsol1}) corresponds to a CDM dominated Universe and $ii)$ from (\ref{cs2alphainf}) for $x > 1$ the speed of sound is exponentially vanishing for $\alpha \to \infty$ so that we can reasonably assume that $c_{\rm s}^{2} \approx 0$. Therefore, if we substitute \eqreff{gcgH2alphainfsol1} and $c_{\rm s}^{2} = 0$ into \eqreff{eqPhi} we obtain the same evolution equation for the gravitational potential as the one in a CDM dominated Universe, see \cite{Mukhanov:1990me}. In this instance, neglecting the decaying mode, $\Phi' = 0$ and no ISW effect is produced.

Write then \eqreff{eqPhi} combined with \eqreff{gcgH2alphainfsol2} and, from the leading order in $x^{3\alpha}$ in (\ref{cs2alphainf}) for $x < 1$, $c_{\rm s}^{2} = \alpha$:
\begin{equation}\label{eqPhialphainf}
\Phi'' + \frac{3\left(1 + \alpha\right)}{1 + \eta_0 - \eta}\Phi' + \left[\frac{3(1 + \alpha)}{(1 + \eta_0 - \eta)^2} + k^{2}\alpha\right]\Phi = 0\;.
\end{equation}
Defining $y \equiv 1 + \eta_0 - \eta$, we recast \eqreff{eqPhialphainf} in the following form:
\begin{equation}\label{eqPhialphainfrecast}
\ddot{\Phi} - \frac{3\left(1 + \alpha\right)}{y}\dot{\Phi} + \left[\frac{3\left(1 + \alpha\right)}{y^2} + k^{2}\alpha\right]\Phi = 0\;,
\end{equation}
where the dot denotes derivation with respect to $y$. Equation~(\ref{eqPhialphainfrecast}) can be solved exactly in terms of Bessel functions:
\begin{equation}\label{solbessel}
\Phi = y^{\frac{3\alpha}{2} + 2}\left[C_1{\rm J}_{\frac{3\alpha}{2} + 1}\left(k\sqrt{\alpha}y\right) + C_2{\rm Y}_{\frac{3\alpha}{2} + 1}\left(k\sqrt{\alpha}y\right)\right]\;,
\end{equation}
where $C_1$ and $C_2$ are arbitrary integration constants. For large values of the order, the Bessel functions can be asymptotically expanded as follows \cite{AS}:
\begin{equation}
{\rm J}_{\frac{3\alpha}{2} + 1}\left(k\sqrt{\alpha}y\right) = \left(\frac{{\rm e}k\sqrt{\alpha}y}{3\alpha + 2}\right)^{3\alpha/2 + 1}\frac{1}{\sqrt{3\pi\alpha + 2\pi}}\left[1 - \frac{8}{\alpha} + O\left(\frac{1}{\alpha^2}\right)\right]
\end{equation}
and
\begin{equation}
{\rm Y}_{\frac{3\alpha}{2} + 1}\left(k\sqrt{\alpha}y\right) = -\left(\frac{{\rm e}k\sqrt{\alpha}y}{3\alpha + 2}\right)^{-3\alpha/2 - 1}\sqrt{\frac{4}{3\pi\alpha + 2\pi}}\left[1 + \frac{8}{\alpha} + O\left(\frac{1}{\alpha^2}\right)\right]\;.
\end{equation}
To leading order in $1/\alpha$, we plug the above asymptotic expansions into \eqreff{solbessel} and find:
\begin{equation}\label{solbesselalphainf}
\Phi = C_1\left(\frac{{\rm e}k}{3\sqrt{\alpha}}\right)^{3\alpha/2}\frac{y^{3\alpha}}{\sqrt{3\pi\alpha}} - 2C_2\left(\frac{{\rm e}k}{3\sqrt{\alpha}}\right)^{-3\alpha/2}\frac{y}{\sqrt{3\pi\alpha}}\;.
\end{equation}
We assume the following initial conditions on the potential $\Phi$ in $\eta = \eta_{\rm tr}$: $\Phi\left(\eta_{\rm tr},k\right) = \Phi_{\rm tr}(k)$ and $\Phi'\left(\eta_{\rm tr},k\right) = 0$. The reason for this choice is that up to $\eta = \eta_{\rm tr}$ the potential behaves like in a CDM dominated Universe, i.e. it is constant. Solution (\ref{solbesselalphainf}) then becomes:
\begin{equation}\label{solbesselalphainf2}
\frac{\Phi}{\Phi_{\rm tr}(k)} = \frac{1}{1 - 3\alpha}\left(\frac{y}{1 + z_{\rm tr}}\right)^{3\alpha} - \frac{3\alpha}{1 - 3\alpha}\frac{y}{1 + z_{\rm tr}}\;.
\end{equation}
For the calculation of the integrals (\ref{ClISW}) and (\ref{ThetalISW}) consider only the contribution proportional to $y$, i.e. 
\begin{equation}\label{solbesselalphainf2approx}
\frac{\Phi}{\Phi_{\rm tr}(k)} \approx \frac{y}{1 + z_{\rm tr}}\;,
\end{equation}
since it is the dominant one for $\alpha \to \infty$. Moreover, assume that the primordial power spectrum $\Delta_{\rm R}^2$ is the Harrison-Zel'dovich scale invariant one and that it propagates invariated up to $\eta_{\rm tr}$. For convenience, define
\begin{equation}
D \equiv \frac{k^3\left|\Phi_{\rm tr}(k)\right|^2}{2\pi^2} = \frac{9}{25}\Delta_{\rm R}^2\;;
\end{equation}
combining the integrals (\ref{ClISW}) and (\ref{ThetalISW}) with the approximated solution (\ref{solbesselalphainf2approx}) and changing the integration variable from the conformal time to the redshift we write:
\begin{equation}\label{ClISWalphainf}
\frac{l(l + 1)C_l^{\rm ISW}}{2\pi D} = \frac{8l(l + 1)}{\left(1 + z_{\rm tr}\right)^2}\int_{0}^{\infty}\frac{\di k}{k}\left[\int_{0}^{z_{\rm tr}}\di z\;{\rm j}_{l}(kz)\right]^2\;.
\end{equation}
Taking into account that ${\rm j}_{l}(kz) = \sqrt{\tfrac{\pi}{2kz}}\;{\rm J}_{l + 1/2}(kz)$, we write \eqreff{ClISWalphainf} as follows:
\begin{equation}\label{ClISWalphainf2}
\frac{l(l + 1)C_l^{\rm ISW}}{2\pi D} = \frac{4\pi l(l + 1)}{\left(1 + z_{\rm tr}\right)^2}\int_{0}^{\infty}\frac{\di k}{k^2}\int_{0}^{z_{\rm tr}}\frac{\di u}{\sqrt{u}}\int_{0}^{z_{\rm tr}}\frac{\di v}{\sqrt{v}}\;{\rm J}_{l + 1/2}(ku){\rm J}_{l + 1/2}(kv)\;.
\end{equation}
Consider now the following case of the Weber-Schafheitlin type integrals \cite{AS}:
\begin{equation}\label{WSformula}
\int_{0}^{\infty}\di t\;\frac{{\rm J}_{l + 1/2}(at){\rm J}_{l + 1/2}(bt)}{t^2} = \frac{1}{4}\frac{b^{l + 1/2}}{a^{l - 1/2}}\frac{\Gamma(l)}{\Gamma(l + 3/2)\Gamma(3/2)}{\rm F}\left(l,-\frac{1}{2};l + \frac{3}{2};\frac{b^2}{a^2}\right)\;,
\end{equation}
where ${\rm F}$ is the Gauss hypergeometric function. Note that formula (\ref{WSformula}) holds true only if $b < a$. We perform the $k$ integration in \eqreff{ClISWalphainf2} according to \eqreff{WSformula} and find the following expression:
\begin{equation}\label{ClISWalphainf3}
\frac{l(l + 1)C_l^{\rm ISW}}{2\pi D} = \frac{4\sqrt{\pi}\;l(l + 1)}{\left(1 + z_{\rm tr}\right)^2}\int_{0}^{z_{\rm tr}}\frac{\di u}{u^l}\int_{0}^{u}\di v\;v^l\frac{\Gamma(l)}{\Gamma(l + 3/2)}{\rm F}\left(l,-\frac{1}{2};l + \frac{3}{2};\frac{v^2}{u^2}\right)\;,
\end{equation}
where we have modified the integration range of $v$ in order to satisfy the condition $v < u$ and thus be allowed to apply \eqreff{WSformula}. Being the integrand function of \eqreff{ClISWalphainf2} symmetric with respect to the line $v = u$, in \eqreff{ClISWalphainf3} we have recovered the correct value of the integral by multiplying by a factor 2.

Expanding ${\rm F}$ in a hypergeometric series and performing the $u$ and $v$ integrations, we find the following series expansion for the ISW contribution:
\begin{equation}\label{ClISWalphainf4}
\frac{l(l + 1)C_l^{\rm ISW}}{2\pi D} = -\frac{l(l + 1)z_{\rm tr}^2}{\left(1 + z_{\rm tr}\right)^2}\sum_{n=0}^{\infty}\frac{\Gamma(l + n)\Gamma(n - 1/2)}{(l + 2n + 1)\Gamma(n + 1)\Gamma(l + n + 3/2)}\;,
\end{equation}
which can be recast in the following more compact form:
\begin{equation}\label{ClISWalphainf5}
\frac{l(l + 1)C_l^{\rm ISW}}{2\pi D} = \frac{2\sqrt{\pi}\;z_{\rm tr}^2}{\left(1 + z_{\rm tr}\right)^2}\frac{\Gamma(l + 1)}{\Gamma(l + 3/2)}\;{}_{3}{\rm F}_{2}\left(l,-\frac{1}{2},\frac{l + 1}{2};l + \frac{3}{2}, \frac{l + 3}{2}; 1\right)\;,
\end{equation}
where ${}_{3}{\rm F}_{2}$ is a generalised hypergeometric function \cite{Erdelyi}.

The asymptotic behaviour of \eqreff{ClISWalphainf5} for large values of $l$ has the following form:
\begin{equation}\label{ClISWalphainfasympt}
\frac{l(l + 1)C_l^{\rm ISW}}{2\pi D} \sim \frac{2\pi z_{\rm tr}^2}{\left(1 + z_{\rm tr}\right)^2}\frac{1}{l}\;,
\end{equation}
which is computed in Appendix~\ref{App}. It can be also directly obtained from the calculations of Kofman and Starobinsky (see Eq.~(12) of \cite{Kofman:1985fp}).

\begin{figure}
\begin{center}
 \includegraphics[width=0.7\columnwidth]{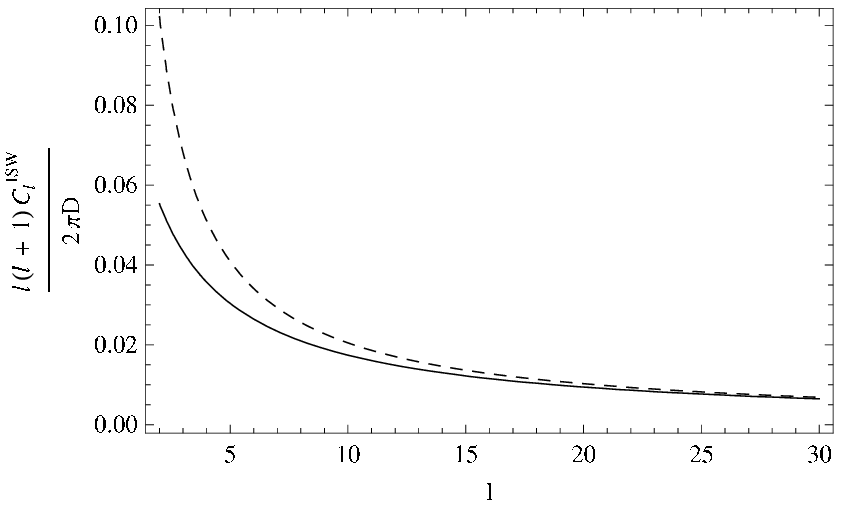}\\
\includegraphics[width=0.7\columnwidth]{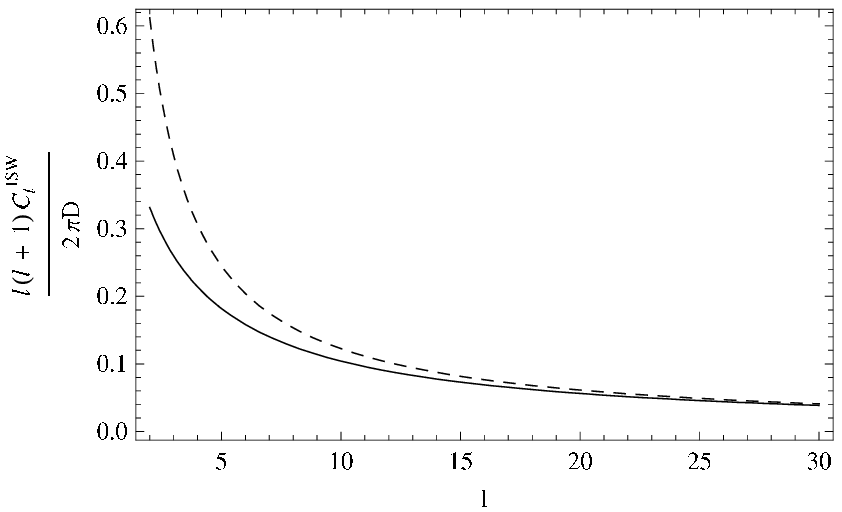}
\caption{Plot of the ISW contribution to the angular power spectrum $l(l + 1)C_l/2\pi$ normalised to $D$ (solid lines) and of its asymptotic form for large $l$'s (dashed lines). Upper panel: $z_{\rm tr} = 0.22$. Lower panel: $z_{\rm tr} = 0.79$.}
\label{FigISW}
\end{center}
\end{figure}

In Fig. \ref{FigISW} we plot \eqreff{ClISWalphainf5} and the relative asymptotic expansion \eqreff{ClISWalphainfasympt} as functions of $l$ for $z_{\rm tr} = 0.22$ and $z_{\rm tr} = 0.79$. The former value is the best fit from the SNIa data analysis performed in the next section.

The black line in the lower panel of Fig. \ref{FigISW} could be associated to those drawn in Fig.~5 of \cite{Urakawa:2009jb}, where the ISW effect contribution is computed for the gCg up to $\alpha = 10$. Note that the $\alpha\to\infty$ contribution is smaller than the $\alpha = 0$ one, which corresponds to the $\Lambda$CDM case and was computed for the first time by Kofman and Starobinsky \cite{Kofman:1985fp}. We expected this result for the following reason. When $\alpha \to \infty$ the Jeans wavenumber diverges. Therefore, according to \cite{Bertacca:2007cv}, the contribution to \eqreff{ClISWalphainf} proportional to the fourth power of the speed of sound does not exist and the behaviour of $l(l + 1)C_l^{\rm ISW}/2\pi$ depends principally on the background evolution. The latter is pretty much similar to the $\Lambda$CDM one. The relevant difference is that for the gCg $\alpha\to\infty$ model the transition from the CDM-like phase to the dS one is very sharp whereas for the $\Lambda$CDM is much smoother. Since in the former case the CDM-like phase lasts longer, the intensity of the ISW effect is lesser.

Finally, note also how in Fig.~5 of \cite{Urakawa:2009jb} the trend of a decreasing ISW effect contribution can be already distinguished for $\alpha = 10$ and large $l$'s.

\section{Background evolution for large $\alpha$ and SNIa data analysis}

In this section we address more quantitatively the behaviour of the background evolution for large values of $\alpha$. Let us consider Friedmann equation (\ref{gcgH2alphainf}) to leading order in $\alpha$:
\begin{equation}\label{Hparam}
\frac{H^{2}}{H_{0}^{2}} \sim \left\{
\begin{array}{cl}
\left(\dfrac{1 + z}{1 + z_{\rm tr}}\right)^{3} & \mbox{   for   } z > z_{\rm tr}\\ \\
1 & \mbox{   for   } z \leq z_{\rm tr}
\end{array}
\right.\;.
\end{equation}
As we pointed out in the previous section, \eqreff{Hparam} mimics the expansion of a pure CDM Universe for $z > z_{\rm tr}$ and the one of a dS Universe for $z \leqslant z_{\rm tr}$. Note that, within this scenario, the present equation of state parameter is $w_0 = -1$, in contrast with the $w_0 \approx -0.7$ of the $\Lambda$CDM model.

We now analyse the background evolution given by \eqreff{Hparam} on the basis of the 157 nearby SNIa of the Constitution set \cite{Hicken:2009dk}. The supernovae data consist in an array of distance moduli $\mu$ defined as:
\begin{equation}\label{mu}
\mu = m - M =  5\log\left(\frac{D_{\rm L}}{{\rm Mpc}}\right) + 25\;,
\end{equation}
where $m$ and $M$ are, respectively, the apparent and the absolute
magnitudes and $D_{\rm L}$ is the luminosity distance expressed in Mpc units:
\begin{equation}\label{d}
D_{\rm L}(z) = c(1 + z)\int_{0}^{z}\frac{\di z'}{H(z')} = \frac{c(1 +
z)}{H_0}\int_{0}^{z}\frac{\di z'}{E(z')}\;,
\end{equation}
where $E(z)$ is the Hubble parameter normalised to $H_0$. 

The integral in \eqreff{d} can be exactly solved for the Hubble parameter given in \eqreff{Hparam}:
\begin{equation}\label{ldist}
\int_{0}^{z}\frac{\di z'}{E(z')} = \left\{
\begin{array}{cl}
\left(3z_{\rm tr} + 2\right) - 2\left(1 + z\right)^{-1/2}\left(1 + z_{\rm tr}\right)^{3/2} & \mbox{   for   } z > z_{tr}\\ \\
z & \mbox{   for   } z \leq z_{tr}
\end{array}
\right.\;.
\end{equation}
Following \cite{Riess:1998cb}, we assume flat priors for $z_{\rm tr}$ and $h$ and assume that the distance moduli are normally distributed. The probability density function (PDF) of the parameters has then the following form \cite{Lupton1993}:
\begin{equation}\label{pdf}
 p\left(z_{\rm tr},h|\mu_o\right) = Ce^{-\chi^2\left(h,z_{\rm tr}\right)/2}\;,
\end{equation}
where $\mu_o$ is the set of the observed distance moduli,
\begin{equation}\label{chisquared}
 \chi^2\left(h,z_{\rm tr}\right) = \sum_{i=1}^n \left[\frac{\mu_{o,i} - 5\log\left(D_{\rm L}/{\rm Mpc}\right) - 25}{\sigma_{\mu_{o,i}}}\right]^2
\end{equation}
and the normalisation constant $C$ has the following form:
\begin{equation}
 \frac{1}{C} = \int\;\di z_{\rm tr}\int\;\di h\;e^{-\chi^2\left(h,z_{\rm tr}\right)/2}\;,
\end{equation}
where the integration ranges over the parameters are, in principle, $h \in (-\infty,\infty)$ and $z_{\rm tr} \in (-1,\infty)$. However, we choose the more reasonable ranges $z_{\rm tr} \in (0,2)$ and $h \in (0.5,0.9)$.

In \eqreff{chisquared} $\sigma_{\mu_{o,i}}$ are the estimated errors in the individual distance moduli, including uncertainties in galaxy redshifts and also taking into account the dispersion in supernova redshifts due to peculiar velocities.

\begin{figure}
\begin{center}
 \includegraphics[width=0.7\columnwidth]{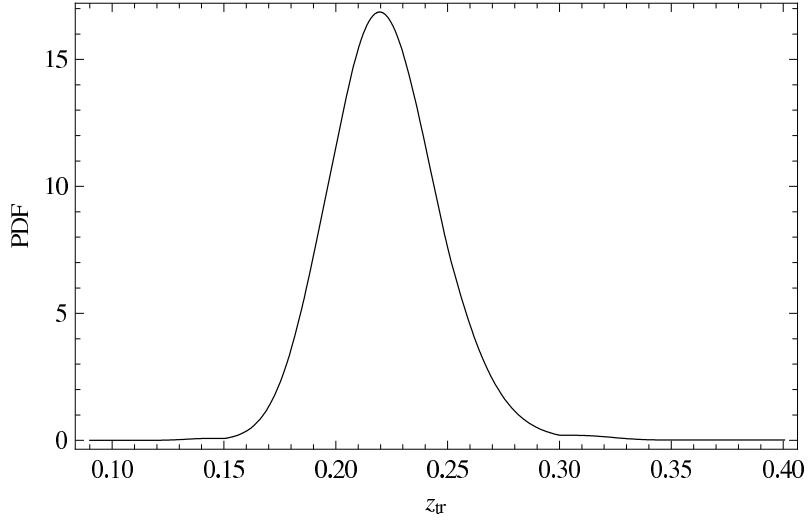}
\caption{Plot of the PDF {\it vs} the parameter $z_{\rm tr}$ after marginalization over $h$.}
\label{Fig2}
\end{center}
\end{figure}

After marginalization over $h$, i.e. integrating \eqreff{pdf} in $h \in (0.5,0.9)$, we show in Fig. \ref{Fig2} the PDF for the parameter $z_{\rm tr}$.
The most probable value is $z_{\rm tr} = 0.222$. At the 68\% confidence level $z_{\rm tr} \in \left(0.198,0.246\right)$, at the 95\% confidence level $z_{\rm tr} \in \left(0.174,0.270\right)$ and at the 99\% confidence level $z_{\rm tr} \in \left(0.154,0.290\right)$. 

\section{Summary and Conclusions}

In the present paper we have investigated the production of ISW effect within the generalised Chaplygin gas cosmological model. Thanks to an argument based on the M\'esz\'aros effect it is possible to find the new constraint $\alpha \gtrsim 350$. For this range of values, the Jeans wavenumber is sufficiently large so that the resulting ISW effect is not strong. Indeed, through a direct calculation, we have found a confirmation of the above qualitative constraint because in the limit $\alpha \to \infty$ the ISW effect contribution to the CMB angular power spectrum is very similar to the one computed for $\alpha = 0$, i.e. for the $\Lambda$CDM model. 

We have then addressed the background evolution of the Universe for $\alpha \to \infty$ and we have found that the model behaves like CDM at early times and then abruptly passes to a dS phase. Taking advantage of the SNIa Constitution set analysis, we have placed the transition at a redshift $z_{\rm tr} = 0.22$. 

In conclusion, it seems that the gCg model has some chances of being viable not only for very small values of $\alpha$ but also for very large ones (we have here limited our discussion to the ISW effect only). However, it must be pointed out that in both cases a degeneracy problem appears: $i )$ for $\alpha \to 0$, it is well-known that the gCg model degenerates into the $\Lambda$CDM; $ii)$ for $\alpha \to \infty$ the degeneration takes place into a ``step-transition'' CDM-dS model. Note that in the second case, the degeneration is not complete. In fact the ``original'' CDM-dS model has a vanishing speed of sound. In the corresponding limit of the gCg, instead, the speed of sound diverges, so the scenario is completely different. A more complete analysis of such picture would therefore be interesting and could perhaps be performed in the framework of the {\it Cuscuton} model introduced in \cite{Afshordi:2006ad, Afshordi:2007yx}.

\acknowledgments{I wish to thank V. Gorini, A. Yu. Kamenshchik, T. Kobayashi, U. Moschella, A.A. Starobinsky and Y. Urakawa for useful comments and suggestions and the Institute of Cosmology and Gravitation (ICG), Portsmouth (UK), for the kind hospitality during the final part of this project. I am indebted with D. Bertacca, S. L. Cacciatori and J. Fabris for invaluable discussions.}

\newpage

\appendix

\section{Calculation of the asymptotic behaviour of the ISW effect contribution for large $l$'s}\label{App}

Consider \eqreff{ClISWalphainf5}, recast in the following way:
\begin{equation}\label{ClISWalphainf5new}
\frac{l(l + 1)C_l^{\rm ISW}}{2\pi D} = \frac{2\sqrt{\pi}z_{\rm tr}^2}{\left(1 + z_{\rm tr}\right)^2}\frac{\Gamma(l + 1)}{\Gamma(l + 3/2)}\;{}_{3}{\rm F}_{2}\left(\frac{l + 1}{2}, l,-\frac{1}{2};l + \frac{3}{2}, \frac{l + 3}{2}; 1\right)\;.
\end{equation}
Adopt the following transformation for the ${}_{3}{\rm F}_{2}$ function \cite{Bailey}:
\begin{equation}\label{3F2trans}
{}_{3}{\rm F}_{2}\left(a, b, c; e, f; 1\right) = \dfrac{\Gamma(e)\Gamma(f)\Gamma(s)}{\Gamma(a)\Gamma(s + b)\Gamma(s + c)}\;{}_{3}{\rm F}_{2}\left(e - a, f - a, s; s + b, s + c; 1\right)\;,
\end{equation}
where $s = e + f - a - b - c$ and $\{a, b, c, e, f\}$ are arbitrary parameters. Employ \eqreff{3F2trans} for the generalised hypergeometric function in \eqreff{ClISWalphainf5new} and find:
\begin{equation}\label{asympt}
\frac{l(l + 1)C_l^{\rm ISW}}{2\pi D} = \dfrac{2\sqrt{\pi}z_{\rm tr}^2}{\left(1 + z_{\rm tr}\right)^2}\dfrac{\Gamma\left(\frac{l + 3}{2}\right)\Gamma\left(l + 1\right)\Gamma(3)}{\Gamma\left(\frac{l + 1}{2}\right)\Gamma(5/2)\Gamma(l + 3)}\;{}_{3}{\rm F}_{2}\left(1, \frac{l}{2} + 1, 3; \frac{5}{2}, l + 3; 1\right)\;.
\end{equation}
Making explicit the series expansion of the generalised hypergeometric function we obtain
\begin{equation}\label{asympt2}
\frac{l(l + 1)C_l^{\rm ISW}}{2\pi D} = \frac{2\sqrt{\pi}z_{\rm tr}^2}{\left(1 + z_{\rm tr}\right)^2}\dfrac{\Gamma\left(\frac{l + 3}{2}\right)\Gamma(l + 1)}{\Gamma\left(\frac{l + 1}{2}\right)\Gamma(\frac{l}{2} + 1)}\sum_{n=0}^{\infty}\dfrac{\Gamma\left(\frac{l}{2} + 1 + n\right)\Gamma(n + 3)}{\Gamma(n + 5/2)\Gamma(l + 3 + n)}\;.
\end{equation}
We approximate the Gamma functions for large $l$'s by means of Stirling formula \cite{AS} and find
\begin{equation}\label{asympt3}
\frac{l(l + 1)C_l^{\rm ISW}}{2\pi D} \sim \frac{\sqrt{\pi}z_{\rm tr}^2}{\left(1 + z_{\rm tr}\right)^2}\;\dfrac{1}{l}\;\sum_{n=0}^{\infty}\dfrac{\Gamma(n + 3)}{2^{n}\Gamma(n + 5/2)}\;.
\end{equation}
The series in \eqreff{asympt3} can be summed:
\begin{equation}\label{seriesPi}
\sum_{n=0}^{\infty}\dfrac{\Gamma(n + 3)}{2^{n}\Gamma(n + 5/2)} = 2\sqrt{\pi}\;,
\end{equation}
and from \eqreff{asympt3} we obtain the result of \eqreff{ClISWalphainfasympt}.

\end{document}